# The Cross-Domain Re-interpretation of Artistic Ideas

**Apara Ranjan (apara.ranjan@ubc.ca), Liane Gabora (liane.gabora@ubc.ca), and Brian O'Connor (brian.oconnor@ubc.ca)**
Department of Psychology, University of British Columbia
Okanagan campus, 3333 University Way
Kelowna BC, V1V 1V7, CANADA

## Abstract

The goal of this study was to investigate the translate-ability of creative works into other domains. We tested whether people were able to recognize which works of art were inspired by which pieces of music. Three expert painters created four paintings, each of which was the artist's interpretation of one of four different pieces of instrumental music. Participants were able to identify which paintings were inspired by which pieces of music at statistically significant above-chance levels. The findings support the hypothesis that creative ideas can exist in an at least somewhat domain-independent state of potentiality and become more well-defined as they are actualized in accordance with the constraints of a particular domain.

## Introduction

Although much social interaction occurs directly through words or actions, a great deal of what humans attempt to communicate, such as ideas for works of art, science, or technology, are not readily expressed through these channels. Complex ideas are therefore often communicated indirectly by way of artifacts. There is evidence that artists leave something of themselves—their own personal signatures or creative styles—in their artifacts, and that creators' identities are recognizably present in their creative works. For example, creative writing students familiar with each other's writing identified significantly above chance, not just which of their creative writing classmates had written each particular piece of writing but which of their creative writing classmates had created each artwork (Gabora, 2010; Gabora, O'Connor, & Ranjan, 2012). Thus, at least in some cases, if a viewer is familiar with the works of creators in a particular domain (such as creative writing), it is possible for the viewer to recognize which creator generated which work, and this is even the case if the works are in a different domain (such as art).

This does not, however, imply that creative artifacts are just the external expression of individual style. We suggest that artifacts constitute a beehive of hidden social interaction, and that their forms reflect, in part, the attempt to transcend one's individuality, *i.e.*, to relinquish oneself to the essence of an idea. We suggest that when personal style is recognizably evident in a work, this is not necessarily due to an attempt to *express* this style, but a side effect of participating in the human enterprise of interactively evolving cultural outputs by adapting them to different personal styles, perspectives, and modalities. The creative process involves not just accessing and combining knowledge, experiences, and ideas, but also inspiration, translation, and re-interpretation (Cropley, 1999; Feldhusen, 1995, 2002; Munford & Gustafson, 1988; Sternberg & Lubart, 1995). Components of a creative work may originate from oneself, others with whom one has communicated directly or indirectly by way of others, or even multiple individuals through the course of history who each put their own spin on it. Inspiration may come from a work in same domain as the work it inspires (as when one poem inspires an idea for another poem). Alternatively, an idea may first be expressed by one individual in one domain, and subsequently translated by someone else into another domain (as when a piece of music gets re-cast in another musical genre, or even inspires a poem). With the advent of new technologies and social media, the distinction between social interaction and individual creative expression becomes increasingly blurred. For example, as one moves from face-to-face communication, to avatar-mediated communication, to music inspired by and intended for someone else, to background music to accompany the activities of a particular cartoon character, to music composed with no obvious inspirational source, it is difficult to draw the line between social interaction and individual self-expression, and cross-modal perception.

The goal of this research was to test the hypothesis that it is possible to recognize the source of inspiration for a creative work when that source of inspiration comes from a different medium. There are several phenomena that suggest that a creative work need not be in the same domain as the inspirational source for the work.

### Related Phenomena

We now review phenomena that point to cross-domain interpretation of ideas as a source of the character of creative works: synesthesia, ekphrastic expression, and cross-domain style.

**Synesthesia** Individuals referred to as synesthetes naturally and spontaneously translate stimuli into another sensory domain. For example, they may see particular letters or numbers in particular colors. Ramachandran (2003) proposed that synesthesia occurs as a result of hyper-connectivity in the brain due to partial collapse of the barrier between sensory domains. Artists, poets, and novelists, are more likely than average to be synesthetes, which suggests that synesthetically driven re-interpretation of inputs from one modality to another can play a role in these creative domains (Ramachandran, 2003).

**Ekphrastic Expression** There is a tradition in the arts of interpreting art from one medium (*e.g.*, oil paint) into another (*e.g.*, watercolor) and thereby coming to know its underlying essence. This practice is referred to as *ekphrastic expression*. The idea behind ekphrastic expression is that an artist may have a more direct impact on an audience by translating art from one medium into another medium because this involves capturing, and thereby becoming intimate with, its underlying form or essence. Ekphrasis may be related to the late nineteenth Century practice of associating particular kinds of music with particular colors. There is anecdotal evidence that music of this time frequently served as a direct inspiration for paintings, and musical terminology was used as titles for paintings. Ekphrastic expression is not just a phenomenon of the past. Modern day film composers attempt to compose music that conveys the emotional tone of the events portrayed in the film, thereby heightening the viewer's experience of these events. Thus film scoring can be seen as a form of ekphrasis. The application of ekphrastic methods in the arts supports the idea that creative individuals extract patterns of information from the constraints of the domains in which they were originally expressed and transform them into other domains.

**Cross-Media Style** Another reason to suspect that the character of creative works arises through cross-domain re-interpretations of ideas is the widespread phenomenon of *cross-media style*. This refers to artistic style that is demonstrated by works of art in more than one medium. For example, the term rococo is applied to a style of painting, sculpture, literature, and music of the 18th Century. Works in a given style are thought to derive from abstract archetypal forms or potentialities that make the artistic mind want to explore different arrangements or manifestations (Burke, 1957).

**Cross-modal Perception** The phenomenon of cross-media style provides evidence that creative works in different media may be similar in terms of psychophysical, collative, and ecological properties (Hasenfus, 1978). Aesthetic perceptions stimulated by creative works may generate emotional, cognitive, behavioral, and/or physiological responses that are amenable to re-expression in another form. This may arise in part due to regularities with respect to the choice of elements (*i.e.*, colors, shapes, words) and/or how they are used (*e.g.*, in an orderly or chaotic manner) (Berlyne, 1971). Studies indicate that there are non-arbitrary mapping between properties of vision and sound (Griscom & Palmer, 2012; Mark 1975, Melara, 1989; Melara, & Marks, 1990; Melara & O'Brien, 1987; Palmer, Langlois, Tsang, Schloss, & Levitin, 2011; Ward *et al.* 2006). For example, the processing of some visual features, such as spatial frequency and lightness, can be affected by auditory features such as pitch and timbre (Mark, 1987).

In the study that perhaps comes closest in spirit to this one, composers were asked to write music inspired by four simple line-drawn shapes: a square, a lightning bold, a curvy shape, and a jagged shape (Willmann, 1944). Music inspired by the same shape was more similar than music inspired by another shape with respect to tempo, melodic pattern, mood, and other characteristics, and listeners could match above chance the music to the shape that inspired it. However, the music could not be said to be reinterpretations of creative works, for the impoverished nature of the stimuli undoubtedly limited the scope for creative expression. The study reported here is the inverse of Willmann's; it investigates not music inspired by art but art inspired by music. Moreover, the goal was to go beyond simple cross-modal mappings to convey in another domain the rich emotionality of genuinely creative works.

## Methods

This study examined whether people were able to correctly recognize which works of art were inspired by which pieces of music. The study was divided into two phases. In the first phase, expert artists created four paintings, each of which was the artist's interpretation of one of four different pieces of instrumental music. In the second phase, naïve participants attempted to determine which piece of music was used as the source of inspiration for each artwork.

### Phase One

**Participants** Two local expert artists, each with approximately 25 years of experience in the field of painting, were recruited for this study. They each received 50$ for their participation.

**Musical Stimuli** Four pieces of piano music from commercially produced sound track CDs with no vocal tract and no other instrumentation were used as stimuli to inspire art. They were selected from a pool of 45 pieces chosen as exemplary of different musical styles: baroque classical, romantic, jazz, and contemporary. Each of these original 45 pieces of music was cropped to three minutes duration, and then rated by three raters on 64 descriptive adjectives on five point Likert scales. The adjectives were derived from previous research on the collative properties of stimulus patterns, specifically, measures of affective reactions to artwork (Berlyne, 1974), and the affective circumplex (Russel, 1980; Watson & Tellegen, 1985). The raters had no previous musical training.

Factor analysis and multidimensional scaling were used to compute the basic dimensions of aesthetic experience in the ratings, and to reveal how the 45 pieces of music were dispersed in the dimensional spaces. The Euclidean distances between the pieces of music across the spaces were used to select four pieces of music from different regions that were clearly dissimilar from each other. The four selected pieces of music were:

(1) 'Love is a Mystery' by Ludovico Einaudi
(2) Number 29 B Flat Major', by Ludwig van Beethoven
(3) 'Circus Gallop' by Marc-André Hamelin
(4) 'All of Me' by Jon Schmidt

**Creation of Artworks** Each of the two artists created one painting for each of the four pieces of piano music, for a total of 8 paintings. On days that paintings were to be created, each artist was provided with a single piece of music and asked to reinterpret it as a painting, *i.e.*, to paint what the music would look like if it were a painting. They were instructed to paint while listening to the music, and encouraged to listen to it as many times as they wished while they painted. They were allowed to use whatever painting supplies they thought could most effectively express the music (*e.g*., watercolors, oils, and acrylics were all acceptable). They were instructed to complete their paintings in one sitting without interruption. They were instructed to take up to a maximum of 120 minutes to listen to the music and complete the painting. The paintings were created in the artists' personal studios. In order to limit the influence of the previous pieces of music on the new painting, the artists were instructed not to re-listen to the piece of music after the painting was finished, and there was a gap of four days between each painting session.

Representative examples of the music-inspired paintings obtained in Phase One of the study are provided in Figures 1, 2 and 3. These paintings constituted the stimuli that were used in Phase Two. Figures 1 and 2, painted by *the same* artist in response to *different* pieces of music, provide the reader with a qualitative sense of the extent to which an artist's personal style comes through in different paintings.

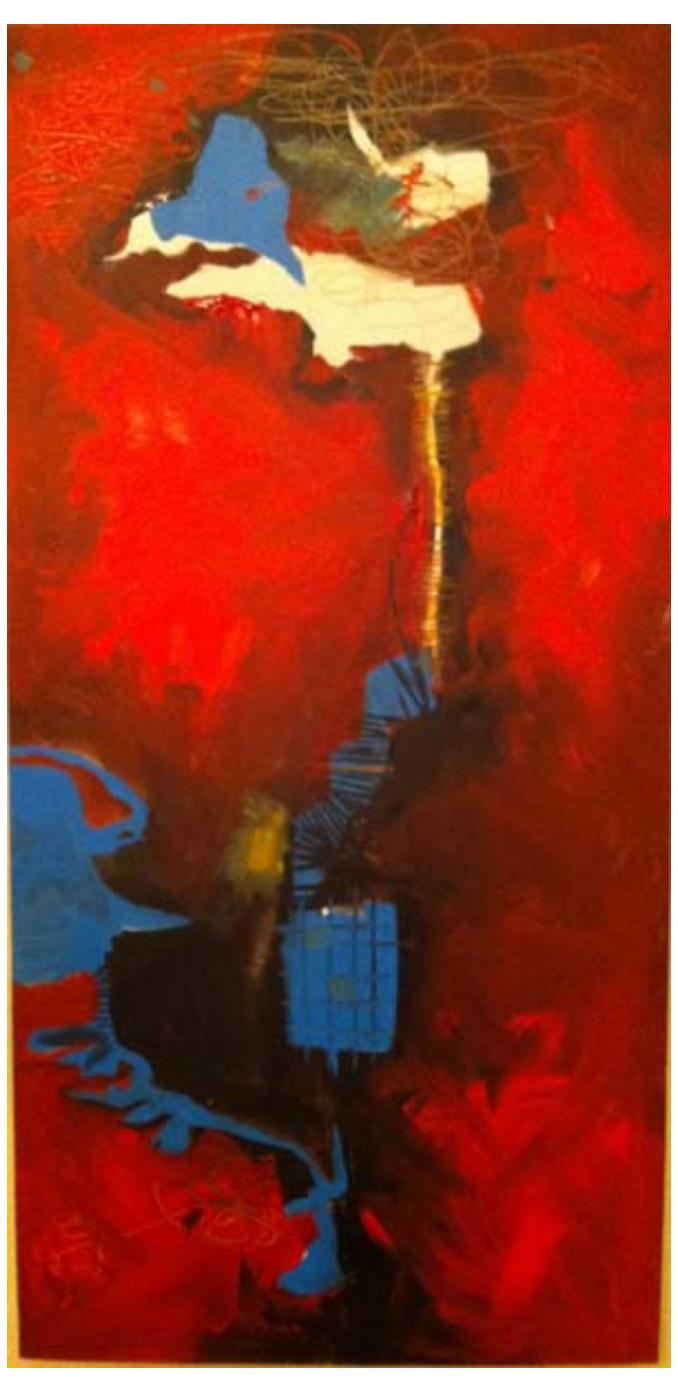

Figure 1. A painting generated by first of the artists as an interpretation of the piece Number 29 B Flat Major', by Ludwig van Beethoven.

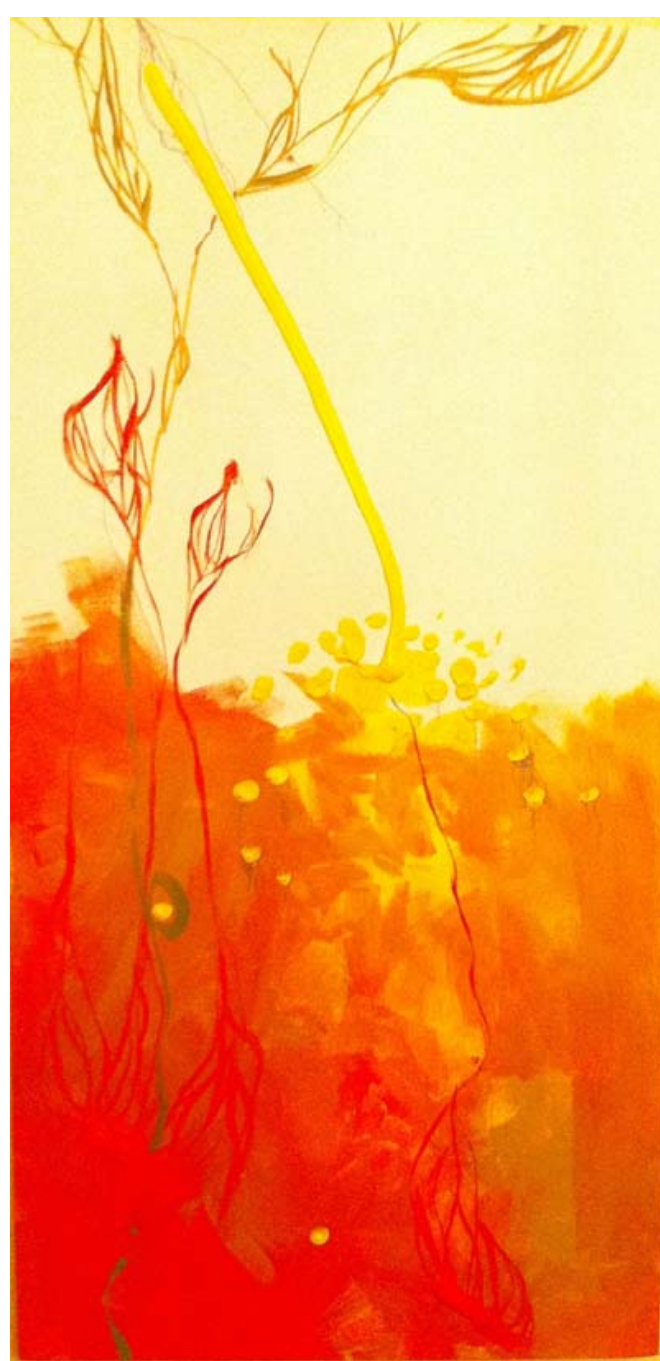

Figure 2. A painting generated by first artist as an interpretation of the piece 'All of Me' by Jon Schmidt.

By comparing figures 2 and 3, painted by *different* artists in response to the *same* piece of music, the reader can obtain a qualitative sense of how a common musical source of inspiration manifests in different paintings.

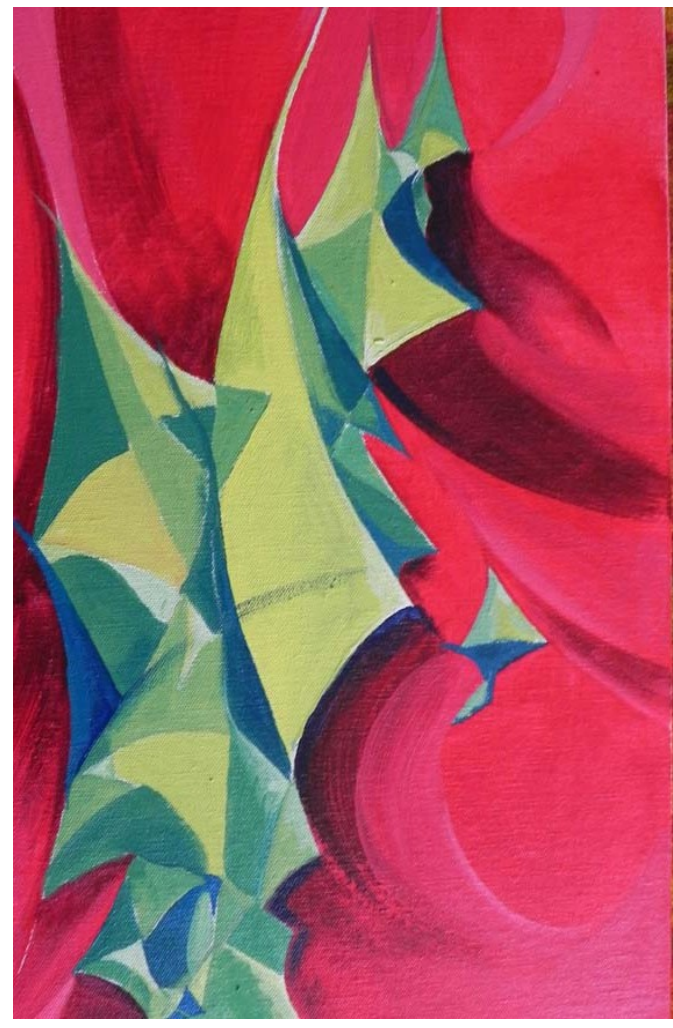

Figure 3. A painting generated by the second artist as an interpretation of the piece 'All of Me' by Jon Schmidt.

## Phase Two

In the second phase of the study we tested whether it was possible to recognize which pieces of music were interpreted as which paintings.

**Participants** The participants were two groups of undergraduates enrolled in psychology courses at the University of British Columbia, consisting of 107 and 89 students respectively, for a total of 196 students. They received partial course credit for their participation.

**Analytic Methods** Two statistics, *H* and *Hu*, were computed to assess the accuracies of the participants' paintings-to-music matches. *H* is the simple hit rate, or the proportion of correct guesses. *Hu* comes from signal detection theory (Wagner, 1993). It corrects for chance guessing and for response bias, such as the tendency to use particular response categories more or less than other response categories. For each set of paintings (i.e., for paintings by artist one and artist two), two hit rate statistics were computed for each participant. One-sample t-tests and a data randomization procedure (Manly, 2007) were then used to assess the statistical significance of the mean *H* and *Hu* values. The tests indicated whether the mean *H* and *Hu* values were significantly different from the *H* and *Hu* values that would have been obtained had participants provided random guesses.

**Procedure and Materials** This part of the study was set up online. There were two sets of the study, one for each artist. In each set, there were the four pieces of music and the four paintings created in phase one by each artist. Each painting was displayed on a web page. Next to each painting were links to the four pieces of music. Two groups of participants consisting of 89 and 107 students were asked to look at the painting and to listen to the four pieces of music respectively. They were asked to identify which piece of music inspired each painting.

## Results

The results are summarized in Table 1.

Table 1: Mean hit rates, *t*-test values, and *r* effect size for identification of paintings inspired by pieces of music. All hit rates and t values were statistically significant.

| | Mean Hit Rate | Chance Hit Rate | *t(df)* | *r* Effect Size |
|---|---|---|---|---|
| Artist One Hit Rate (*H*) | .35 | .24 | 4.0 (88) | .39 |
| Artist One Unbiased Hit Rate (*Hu*) | .36 | .24 | 4.0 (88) | .40 |
| Artist Two Hit Rate (*H*) | .44 | .25 | 6.3(106) | .52 |
| Artist Two Unbiased Hit Rate (*Hu*) | .46 | .25 | 6.3(106) | .52 |

For the first artist, the mean hit rates were $H = .35$ and $Hu = .36$. The mean hit rates that would have been obtained on the basis of random guesses for these questions were .25 and .25, respectively. Both hit rates are statistically significant according to both conventional and data randomization t-tests ($t(88) = 4.0$, $t(88) = 4.0$, $p < .001$), and the *r* effect sizes were large, .39 and .40. For the second artist, the mean hit rates were high, $H = .44$ and $Hu = .46$, statistically significant according to both conventional and data randomization t-tests ($t(106) = 6.3$, $t(106) = 6.3$, $p < .001$), and the effect sizes were large, $r = .52$ and $r = .52$. Thus participants identified at above-chance levels which paintings were inspired by which pieces of music for both artist one and artist two.

## Discussion

There is a longstanding debate concerning the extent to which the semantic complexity of artistic works is amenable to scientific methods (Becker, 1982). We tested the hypothesis that the core idea behind a creative work is recognizable when it is translated from one domain to another. To our knowledge, the only other previous study to test this hypothesis (Willmann, 1944) used highly artificial stimuli that most would not consider creative works in and of themselves. The hypothesis was supported by our finding that when pieces of music were re-interpreted as paintings, naïve participants were able to guess significantly above chance which piece of music inspired which painting. Although the medium of expression is different, something of its essence remains sufficiently intact for an observer to detect a resemblance between the new work and the source that inspired it. The results are consistent with a number of phenomena familiar to artists, mentioned in the Introduction, namely synesthesia, ekphrastic expression, cross-media style, and cross-modal perception.

The research reported on here may be a step toward distinguishing between domain-specific and domain-general aspects of creative works. We suggest that at their core, creative ideas may be much less domain-dependent than they are generally assumed to be. Our results support the view that the uniqueness of a creative work derives at least in part from, not just the personal style of the creator, but from encounters with works in domains that differ from the domain of the creative output, or even different kinds of experiences altogether. In other words, it is possible for the domain-specific aspects to be stripped away such that the creative work exists in an abstracted state of potentiality at which point they are amenable to re-expression in another form. A creative idea may exist in form that is freed of the constraints of a particular domain, and that the creator's job may be in part to, to simply allow that domain-independent entity to take a particular form, using domain-specific expertise and the tools of his or her trade. Over time they may become more fully actualized, and well-defined, as they are considered from different perspectives in accordance with the constraints of the domain in which they are expressed.

The capacity for cross-domain translation of creative ideas supports the hypothesis that an individual's creative outputs are expressions of a particular underlying uniquely structured self-organizing internal model of the world, or worldview. Our creative abilities may be a reflection of the tendency of a worldview to transform in such a way as to find connections, reduce dissonance, and achieve a more stable structure (Gabora & Merrifield, 2012). This view of creativity is consistent with previous research showing that midway through a creative process, an idea may exist in a 'half-baked' state of potentiality, in which one or more elements are ill-defined (Gabora, 2005, Gabora & Saab, 2011). When a work is translated from one domain (*e.g*., music) into another (*e.g*., painting), the two works may be recognizably related because the process by which the worldview assimilates or comes to terms with the works is at a structural level deeply analogous.

Although that idea that at least some creative tasks involve the abstraction and re-expression of 'raw' potentialities or forms seems obvious to many artists we have spoken with, it stands in contrast with most academic theories of creativity. Creativity is typically portrayed as a process of searching and selecting amongst candidate ideas that exist in discrete, well-defined states. This can be traced back to early views arising in the artificial intelligence community, wherein creativity was thought to proceed by heuristically guided search through a space of possible solutions (Newell, Shaw & Simon, 1957; Newell & Simon, 1972; Simon, 1973, 1986) or possible problem representations (*e.g.,* Kaplan & Simon, 1990, Ohlsson, 1992). The view that creativity proceeds through a process of search and selection is also assumed in more contemporary theories, such as the theory that creativity is a Darwinian process; *i.e.*, new ideas are obtained by generating variations more or less at random and selecting the best (*e.g.,* Campbell, 1965; Simonton 1999a,b, 2005).

Our results bring up the question of what it was about the paintings that made it possible to trace them to the artworks that had inspired them. We are currently investigating to what extent people assign similar experience variable ratings and similarity ratings to paintings and the music that inspired them and whether these ratings correspond even when participants cannot *identify* which piece of music inspired the painting. A possible clue to the mechanisms underlying cross-domain interpretation of creative ideas comes from research by Feedberg and Gallese (2007) on perceiving action in artwork. They propose that art observers implicitly imitate the creative actions undertaken by the artist in the making of the work. In our study it is possible that observers were not just perceiving action in art but were also able to match qualities of the art with qualities such as the rhythm and tempo of the music that inspired it. The phenomenon of action perception in paintings could also at least partially account for the ability to recognize the essence of ideas interpreted across domains. In order to recognize the inspiration of an artwork or a cross-media style, expertise in a domain might stimulate the action system while the observer imagines how the artwork was created. Thus, future research will also investigate the role of expertise in the recognition of a connection between works in different domains. We hypothesis that expertise in a domain might increase the activation of the action system while the observer imagines how the artwork got created, thereby enhancing the capacity for recognition of cross-domain re-interpretation in a task such as this.

The effect of inspirational source on creative output may be weaker than the effect of personal style reported earlier (Gabora, 2010; Gabora, O'Connor & Ranjan, 2012), given that paintings by different artists inspired by the same piece of music could be quite different, as seen by comparing Figures 2 and 3. This could however reflect individual differences with respect to which elements of the source had sufficient personal relevance to serve as departure points for the artists' own creative works. This interpretation is consistent with the finding that when pictures were used as stimuli for poetry, poets focused on particular portions of the pictures to serve as the basis for their poems, and different poets focused on different portions (Patrick, 1935). We are currently investigating how these two factors interact, *i.e.,* whether artists' individual styles influence the ease of identifying which music inspired their paintings. Our aim is not to partition out how much creative works owe their distinctive character to their creators and how much they owe to other sources. We suspect that such a partitioning is not possible, that in the most successful creative works there is a fusion of the two, and that the ability to fuse ones' personal style with the inspirational source for a work plays a role in artistic genius. Though commonly portrayed as introverted and withdrawn, the creative genius may, through the assimilation and generation of creative artifacts, be deeply immersed in a form of social interaction that connects all of humanity to the deepest and most influential thinkers our world has known. We suggest that the extent to which the arts feed on the cross-domain adaptation and reinterpretation of ideas has been underappreciated, and that it may in fact play a pivotal role in the evolution of human culture.

## Acknowledgments

We are grateful for funding to the second author from the Natural Sciences and Engineering Research Council of Canada and the Concerted Research Program of the Flemish Government of Belgium. We thank Jon Corbett for assistance with this research.